\documentclass[twocolumn]{aastex63}
\usepackage{epsfig}
\usepackage{natbib}
\usepackage{import}
\usepackage{todo}

\let\tablenum\relax

\usepackage{siunitx} 
\newcommand{\y}{\:{\rm y}}

\newcommand\kms{\:\rm{\,km\,s^{-1}}}

\newcommand\masy{\:\rm{\,mas\:yr^{-1}}}

\newcommand{\chisq}{\chi^2}

\newcommand{\ie}{{\it i.e.}}
\newcommand{\eg}{{\it e.g.}}

\newcommand{\rx}{RX J0822$-$4300}
\newcommand{\pu}{Puppis A}

\shorttitle{{The proper motion of \rx}}
\shortauthors{Martin Mayer, Werner Becker, Daniel Patnaude, P.~Frank Winkler, Ralph Kraft}

\submitjournal{ApJ} 
\accepted{June 28, 2020}

\begin{document}

\title{{The Proper Motion of the Central Compact Object \rx \linebreak  in the Supernova Remnant \pu, Revisited}}
\author[0000-0002-9771-9841]{Martin Mayer}
\affiliation{Max-Planck Institut f\"ur extraterrestrische Physik, Giessenbachstrasse, 85741 Garching, Germany}

\author[0000-0003-1173-6964]{Werner Becker}
\affiliation{Max-Planck Institut f\"ur extraterrestrische Physik, Giessenbachstrasse, 85741 Garching, Germany}
\affiliation{Max-Planck Institut f\"ur Radioastronomie, Auf dem H\"ugel 69, 53121 Bonn, Germany}

\author{Daniel Patnaude}
\affiliation{Smithsonian Astrophysical Observatory, Cambridge, MA 02138, USA}

\author[0000-0001-6311-277X]{P. Frank Winkler}
\affiliation{Department of Physics, Middlebury College, Middlebury, VT 05753, USA}

\author{Ralph Kraft}
\affiliation{Smithsonian Astrophysical Observatory, Cambridge, MA 02138, USA}

\begin{abstract}
\noindent
We present an improved proper motion measurement of the central compact object \rx, located in the supernova remnant Puppis A. By employing a new data set taken in February 2019 by the High Resolution Camera aboard the {\em Chandra} X-ray Observatory, we approximately double the available temporal baseline for our analysis to slightly more than 19 years (7000 days). We correct for the astrometric inaccuracy of {\em Chandra} using calibrator stars with known optical positions that are detected in all observations. Thereby, we obtain absolute positions of \rx\, accurate to around \SI{0.1}{\arcsecond} and from these a new best estimate for its total proper motion of $\mu_{\rm tot}= (80.4 \pm 7.7) \masy$. For a remnant distance of 2 kpc, this corresponds to a projected kick velocity of $(763 \pm 73) \kms $ at a position angle of $\phi_0 = (247.8 \pm 4.4)^{\circ}$. The proper motion measurement of \rx~is used for discussing the kinematic age of Puppis A.  
\end{abstract}


\section{Introduction \label{intro}}
\noindent After their death in violent core-collapse supernovae (\ie~types Ib, Ic, II), massive stars leave behind compact remnants such as black holes or neutron stars. The latter constitute an opportunity to directly observe matter under some of the most extreme conditions in the universe. Over the years, observations have revealed a diverse ``zoo'' of neutron stars: While most young neutron stars are detected as non-thermal pulsed sources in the radio, optical, X- or $\gamma$-ray regime \citep[for an overview, see \eg][]{Becker2009, NSZoo},
members of the class of Central Compact Objects (CCOs) are seen exclusively as isolated hot, steady thermal emitters in X-rays, located at (or near) the center of supernova remnants (SNRs), but without characteristic pulsar wind nebulae. In particular, CCOs have been found associated with relatively young, oxygen-rich Galactic remnants of core-collapse supernovae such as Cas A \citep{tananbaum99,chakrabarty01} and Puppis A \citep{petre96}.
In total, the sample of Galactic SNRs with confirmed CCOs consists of around 10 objects, including G266.1$-$1.2, PKS 1209$-$51$/$52, G330.2$+$1.0, G347.3$-$0.5, G350.1$-$0.3, Kes 79, G353.6$-$0.7 and G15.9$+$0.2 \citep[see][and references therein]{DeLuca17}\footnote{See also \url{http://www.iasf-milano.inaf.it/~deluca/cco/main.htm}}. 

While X-ray pulsations of CCOs have been detected in a few cases, these can be fully explained by the rotational modulation of thermal emission from hotspots on the neutron star surface \citep[e.g.][]{gotthelf10} and are therefore not comparable to strong pulsations across the electromagnetic spectrum seen from typical young pulsars. The exact reason for the existence of these hotspots is still unclear since heating through accretion or particle bombardment seems unlikely \citep{DeLuca17}. 
The lack of nonthermal emission from CCOs can likely be attributed to their comparatively weak magnetic field, which is inferred from small spin-down rates, justifying their designation as ``anti-magnetars" \citep{Gotthelf13}.  
An issue with the description of CCOs as a homogeneous class is the paucity of their descendants, since one would expect to find many more ``orphaned'' compact objects \citep{Halpern10} without visible SNRs in a similar region of the $P-\dot{P}$ parameter space than are observed in practice \citep{Kaspi10}. 

The dynamical imprint of a violent supernova explosion on its remnant can be studied by observing the kinematics of the ejecta, \eg, fast-moving optical filaments, but also by studying the proper motion of the neutron star itself. Typically, contemporary simulations of core-collapse supernovae predict significant explosion asymmetries, which manifest themselves as bipolar jets, large-scale anisotropies,  and/or strong natal kicks to the compact object. These birth kicks can be made plausible simply by conservation of momentum: If a large ejecta mass is expelled at high velocity preferentially in a certain direction, one would naturally expect the compact remnant to experience recoil in the opposite direction. 

After it was found from optical observations that ejecta in Puppis A generally expand towards north and east \citep{Winkler85,Winkler88}, it was expected that a possible compact remnant would be moving towards the southwest. Following the discovery of the CCO \rx\, by \citet{petre96}, the measurement of its kinematics thus became very interesting, even though it is  challenging to achieve sufficient astrometric accuracy for an object that emits exclusively in X-rays. Such a study  became possible only after the launch of {\em Chandra} owing to its unparalleled spatial resolution, once a sufficiently long temporal baseline of 5.3 years (December 1999 - April 2005) between observations had been acquired. 

Two studies successfully measured a proper motion towards the southwest, with marginally consistent results for its absolute value: \citet{HuiBeck06} found $\mu_{\rm tot}= (107 \pm 34) \masy$, while \citet{Winkler07} measured an even larger value of $\mu_{\rm tot}= (165 \pm 25) \masy$. Combined with an approximate distance to the SNR of 2 kpc, the measurement by \citet{Winkler07} implied a very high transverse velocity on the order of $1600\kms$, leading to the designation of \rx\, as a ``cosmic cannonball."\footnote{In the analysis of \citet{Becker12}, it was found that this high velocity was attributable to a subtle bug in the {\em Chandra} software for fitting and locating the off-axis reference stars. Cf.~Section \ref{PSFMod} of this paper.}
While constituting an exciting result on its own, this finding was in conflict not only with observations of pulsar birth velocities being on the order of $\sim 500 \kms$ \citep{caraveo93, frail94, hobbs05}, but also with theoretical predictions from simulations of supernova explosions. However, \citet{Becker12} repeated the above study including a new {\em Chandra} HRC observation taken in August 2010, and found a more conservative value of $\mu_{\rm tot}= (71 \pm 12) \masy$ which corresponds to a velocity of $(672 \pm 115) \kms $, a result in better agreement with theory and the general distribution of measured pulsar proper motion velocities.  

In order to finally ``pin down" the proper motion of \rx\, in direction and magnitude, this work incorporates a new {\em Chandra} observation from early 2019, almost doubling the previously available time baseline. In Section \ref{obs}, we give an overview of the data set we used and the initial data processing. In Section \ref{Analysis}, we describe the analysis steps we used to  obtain the CCO proper motion value from our data. We then discuss the implications on neutron star kick velocity and remnant age in Section \ref{Discussion} and summarize our findings in Section \ref{Summary}.     

\section{Observations and Data Reduction\label{obs}}
\noindent In total, \rx~ has been observed five times with {\em Chandra}'s high resolution camera (HRC). Four of these observations were taken with the HRC-I detector (optimized for imaging), and one with the HRC-S (optimized for spectroscopic readout). In order to reduce the influence of possible small but relevant systematic deviations between the detectors (\eg~due to differences in the degap correction or slight misalignments of the detector axes), we exclude the HRC-S observation from our analysis, leaving four observations spanning $19.18$ years. These consist of three archival observations, and a new one carried out on February 2, 2019. 
A journal of the relevant observation IDs, dates, and observation length is given in Table \ref{table1}.

\begin{deluxetable*}{crccc}[t!] %Or ccccc?
\tabletypesize{\small}
\tablewidth{0pc}
\renewcommand{\arraystretch}{1.15}
\tablecaption{{\em Chandra} observations of \rx\,\label{table1}}
\tablehead{
 Instrument &   ObsID &    Date           & OnTime (s) & Exposure Time (s)}
\startdata        
   HRC-I    &    749  &  1999 Dec 21/22 & 18014  &    \phn9860        \\ 
   HRC-I    &   4612  &   2005 Apr 25   & 40165  &    21317        \\  
   HRC-I    &   11819&  2010 Aug 10/11 & 33681  &    15467        \\
   HRC-I    &   12201&  2010 Aug 11 & 38681  &    17808        \\
   HRC-I    &   20741 &  2019 Feb 02 & 40175  &   19790        \\
\enddata                                                                                                \tablecomments{The 2010 observation was carried out as two consecutive ObsIDs (11819 \& 12201), without intervening repointing \citep[][]{Becker12}. Therefore, we merged the two event files after the reprocessing step using {\tt dmmerge}.}             
\end{deluxetable*}                                                                                      

We acquired the archival observations from the {\em Chandra} Data Archive\footnote{\url{https://cda.harvard.edu/chaser/mainEntry.do}} and reprocessed the data using the standard {\em CIAO} (Chandra Interactive Analysis of Observations \citep{CIAORef}) script {\tt chandra\_repro} to create new level 2 event files on which we base our analysis. In this and all subsequent steps we used {\em CIAO} version 4.9 and {\em CALDB} version 4.7.4. We checked the observations for flares by inspecting the light curves of point-source-free regions. For the latest observation (ID 20741), we found a background flare affecting the data during around 8\% of the time. Therefore, we excluded the affected time intervals for an effective exposure of $18195\,\si{s}$, yielding a significantly reduced particle background while hardly affecting the number of source counts.  

\begin{figure}[h!]
\centering
\includegraphics[width=1.0\linewidth]{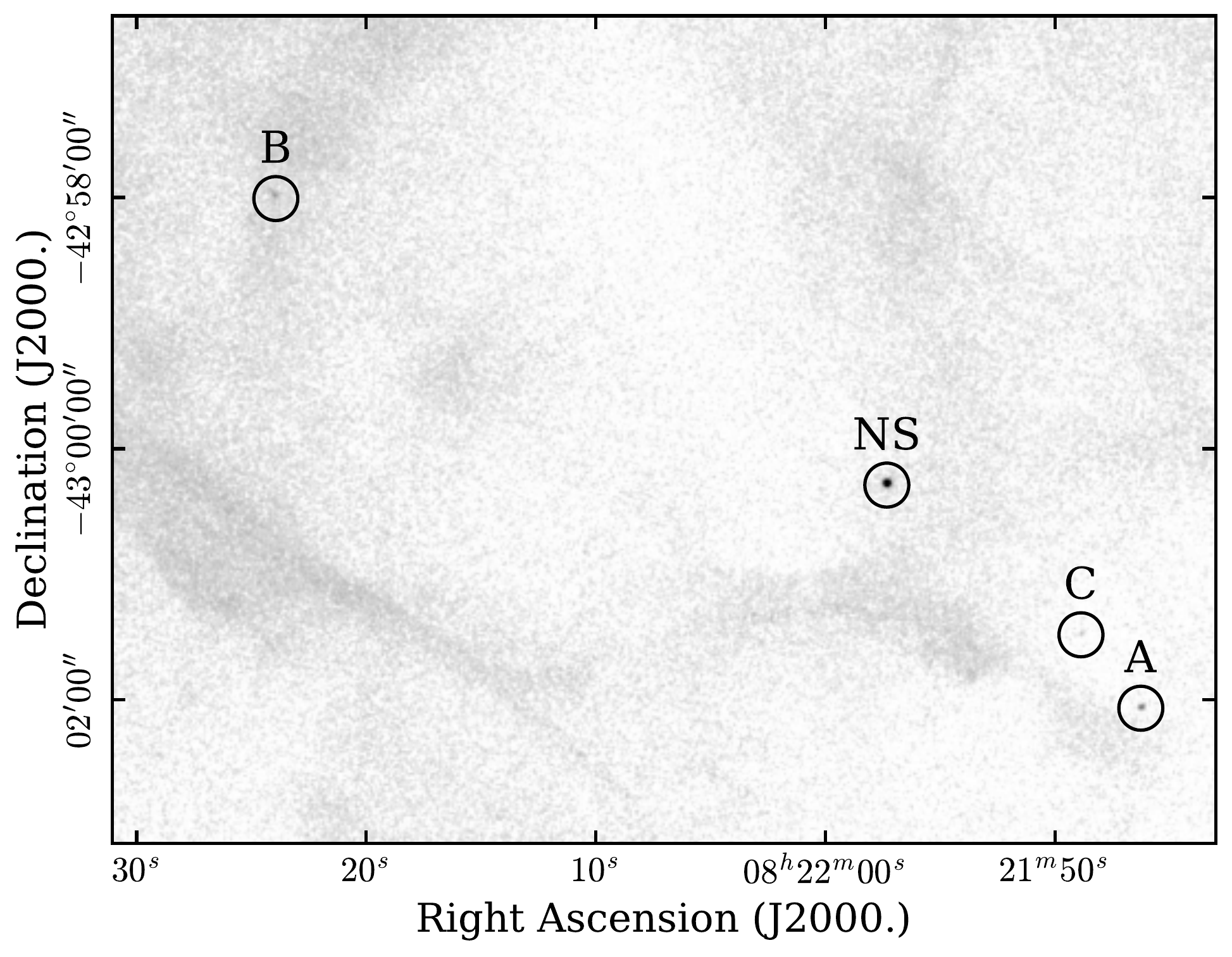}
\caption{Image of the observation from 2010 with the positions of \rx~ and the astrometric calibrator stars indicated. The scale is logarithmic and the image has been slightly smoothed with a Gaussian of width $0.5''$. The field measures around $8.8'\times6.6'$.}
\label{StarOverview}
\end{figure}

Based on the previous results, we can see that we need to achieve an absolute astrometric accuracy significantly below the arcsecond level in order to obtain sufficient precision on our measurement of the proper motion of the CCO. Therefore, while the absolute positional accuracy of {\em Chandra} is unparalleled for X-ray telescopes at $\sim 0.6''$\footnote{\url{http://cxc.harvard.edu/cal/ASPECT/celmon/}}, we still have to improve on the raw astrometric position by a factor of a few in order to obtain a clean signal for the neutron star motion. Our method closely follows that of \citet{Becker12},  albeit with some changes which we will highlight in the next sections. As in the previous works on this topic, we use three nearby optical calibrator stars (designated as A, B, and C) that are detected in X-rays to obtain a precise reference for the world coordinate system (WCS). We take advantage of the {\em Gaia} DR2 catalog\footnote{\url{https://www.cosmos.esa.int/web/gaia/dr2}} which offers strongly improved precision on the positions and proper motions of our astrometric calibrators \citep{GaiaMission,GaiaSummary} compared to the UCAC3 catalog \citep{Zacharias09}. In Table \ref{GaiaTable}, we list the optical properties of the stars, and we indicate their relative location to \rx\, (designated as NS in the following) in Figure \ref{StarOverview}. 

The X-ray image demonstrates the general difficulty of detecting weak point sources on top of diffuse background emission from the SNR. While being very faint in X-rays, source C is located in a region of relatively low diffuse emission and can therefore reliably be detected despite the low associated count rate. Source B, however, is superimposed on bright diffuse emission, which hampers its precise localization. Furthermore, all three calibrator sources are located well off-axis, where the {\em Chandra} point spread function (PSF) becomes increasingly degraded.  Comparing these limitations in the X-ray regime to the exquisite precision of the optical positions of the stars in the {\em Gaia} catalog, we can infer that the dominant error source will lie in the determination of source positions in X-rays, and not in the input astrometric calibration values.

\begin{deluxetable*}{cccccc}[t!]  
\tabletypesize{\small}
\tablewidth{0pc}
\renewcommand{\arraystretch}{1.15}
\tablecaption{IDs, positions and proper motions of the three reference stars. \label{GaiaTable}}
\tablehead{
\multicolumn{2}{c}{Designation} &\multicolumn{2}{c}{Position (Epoch 2015.5)} & \multicolumn{2}{c}{Proper Motion } \\
 Short   & {\em Gaia} Source ID        & R.A. (ICRS)          & Decl. (ICRS)         & $\mu_{\alpha}$  & $\mu_{\delta}$\\
         &               & (h:m:s)       & (d:m:s)     & (mas yr$^{-1}$) & (mas yr$^{-1}$)}
\startdata
  A & 5526323497671973632 &08:21:46.2788&$-$43:02:03.590& $-11.68\pm 0.03$ &$2.70\pm 0.03 $\\
  B & 5526324631543374464 &08:22:24.0044&$-$42:57:59.261& $-0.35 \pm 0.03$ &$8.49\pm 0.03 $\\
  C & 5526323527726140416 &08:21:48.8067&$-$43:01:28.211& $-51.05\pm 0.04$ &$6.82\pm 0.04 $\\ 
\enddata  
\tablecomments{Data as listed in the {\em Gaia} DR2 catalog \citep{GaiaSummary}. The stated $1\sigma$ uncertainties of Right Ascension \& Declination at the reference epoch are significantly below milliarcsecond-level, and would only become relevant at the sixth \& fifth decimal digit, respectively. The proper motion along the Right Ascension \& Declination axes are labelled as $\mu_{\alpha}$ \& $\mu_{\delta}$, respectively} 
\end{deluxetable*}

As a first approximation, we use {\em CIAO}'s wavelet detection algorithm {\tt wavdetect} to estimate the positions and to verify the detectability of all calibrator sources in each observation. 
We then extract sub-images centered on each source for each observation to reduce computation times in the subsequent steps. In order not to significantly oversample the PSF, we use bin sizes of $1\times1$ pixels for the NS, and $2\times2$ pixels for sources A, B, C. Since the calibrator sources are located well off-axis (where the PSF broadens) and have fewer counts compared to the CCO, we do not lose a significant amount of information with the $2\times2$ pixel binning and we suppress pixel-to-pixel fluctuations for the faint sources.

\section{Spatial Analysis}\label{Analysis}
\noindent In order to measure the proper motion of \rx~from our data, we apply a method similar to that described in \citet{Becker12}: We determine the position of all sources (NS, A, B, C) in each data set by modelling and fitting an appropriate PSF to the data (Section \ref{PSFMod}). Combining the measured positions of the calibrator stars with their optical positions, we determine an optimal transformation from {\em Chandra}'s coordinate system to the WCS, which we then apply to the X-ray position of the CCO. In this step, special care is taken in the propagation of uncertainties from each individual fit to the final corrected position of \rx, effectively reducing the final size of the error contours (Section \ref{Trafo}). 
From the absolute positions of the CCO at four epochs, we then straightforwardly determine a new best-fit estimate for its proper motion  in two dimensions (Section \ref{LinearFit}).

\subsection{PSF Modelling and Fitting\label{PSFMod}}

\noindent In order to get the most reliable estimate for the source positions, we simulate a PSF model for each source in each observation.
This is necessary to get results which are unbiased by \eg~the telescope roll angle and pointing, since the off-axis PSF is very broad and, more importantly, distorted. Therefore, the location of peak flux in the PSF does not necessarily correspond to the actual source location. For our simulations, we used the online {\em Chandra} Ray Tracer tool ({\em ChaRT} \footnote{\url{http://cxc.harvard.edu/chart/index.html}}) which incorporates the best available model of the {\em Chandra} high resolution mirror assembly (HRMA). {\em ChaRT} uses an input source position in combination with the aspect solution of the observation to trace the photon trajectories from the sky through the HRMA up to the detector plane. Key parameters are the assumed source monochromatic energy, which we set to $1.0 \,\si{keV}$ \citep[see][]{Becker12} and the source photon flux, which we set to the maximal value of $1\times10^{-2}\,\si{cm}^{-2}\,\si{s}^{-1}$, in order to minimize the influence of Poisson fluctuations on our PSF model. PSF models with a lower simulated source flux may better  resemble the actual images qualitatively, but quantitative fits performed with those yield underestimated positional errors and show larger systematic fluctuations.
The {\em ChaRT} documentation explicitly discourages artificially scaling up source flux to create more rays. However, this warning applies mostly to simulations where one attempts to model nonlinear detection effects like photon pile-up in CCD cameras. The HRC also experiences effects at high rates, such as count rate nonlinearity and deadtime effects, but these do not impact the imaging quality of the source. In total, we create five statistically independent simulations for each source.  

\begin{figure*}[h!]
\centering
%\captionsetup{width=0.85\linewidth}
\includegraphics[width=0.85\linewidth]{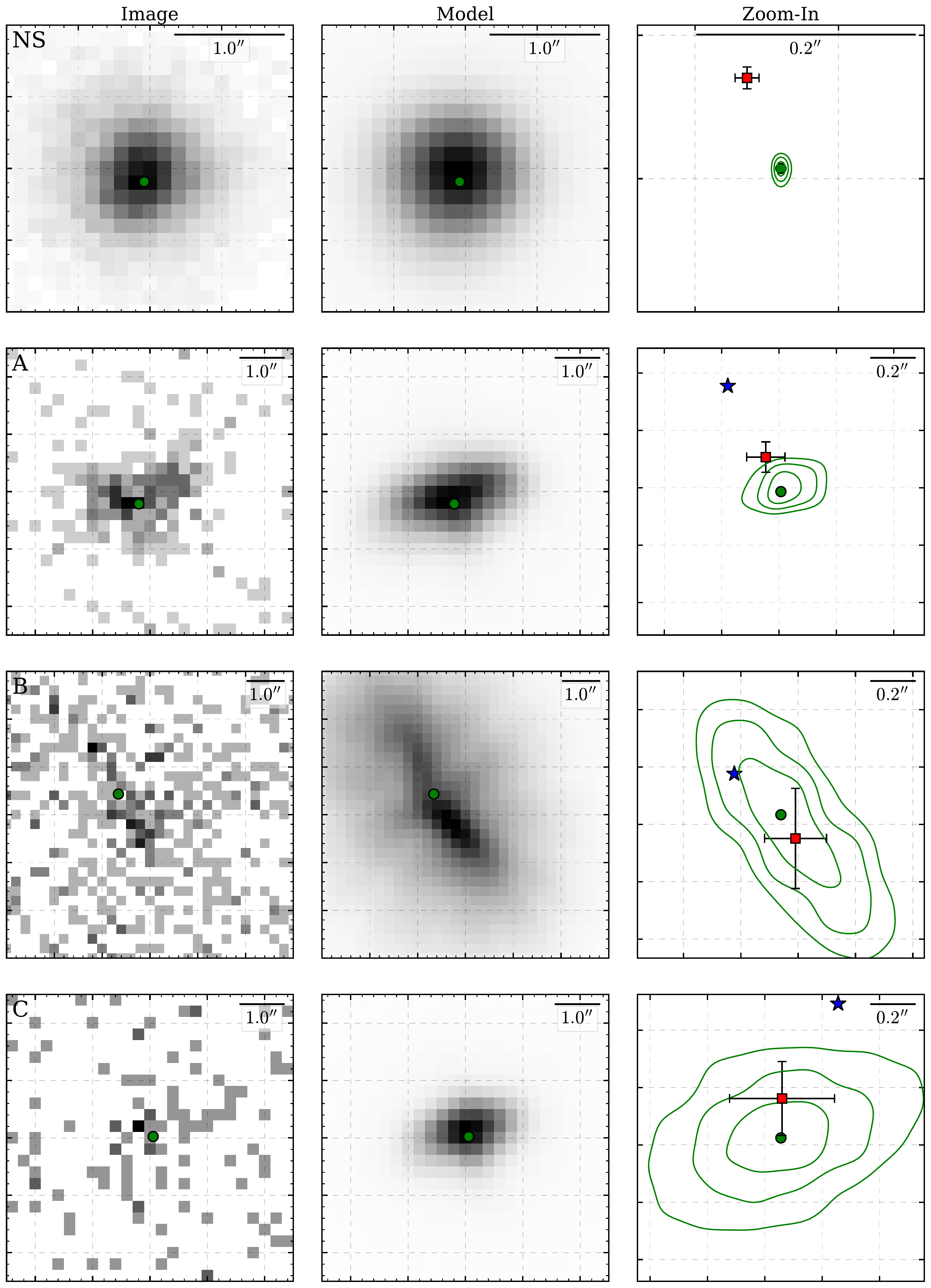}
\caption{Illustration of the PSF fits for the four sources in the 2010 observation (ObsIDs 11819 \& 12201). {\em Left:} Input images; {\em Center:} Best-fit model, \ie~PSF image convolved with best-fit narrow Gaussian; {\em Right:} Zoom-in on the source position. We indicate the best-fit $(x,y)$-position (green circle; as in  Table \ref{FitResults}) with its 1, 2, and 3$\sigma$ uncertainty contours returned by {\tt reg\_proj}; For comparison, we plot the raw best-fit position given in \citet{Becker12} (red square) and the optical position from {\em Gaia} projected on our coordinate system (blue star). Data and model images have been binned by a factor of 2 for sources A, B, C and a square-root intensity scale was used to display them.}
\label{PSFFits}
\end{figure*}

The five ray files created by {\em ChaRT} are then fed simultaneously into {\em MARX} \footnote{\url{http://space.mit.edu/CXC/MARX/}} \citep[][version 5.4.0]{Marx} which we use to project the rays onto the HRC-I detector and to simulate its response to the photons. We finally obtain PSF model images for each source, with the respective binning matching the data. An impression of the morphology of the PSF models is given in the central panel of Figure \ref{PSFFits}. We found that, in order for the PSF to accurately reflect the observed image, we need to set the model detector behavior to ``non-ideal." This applies additional blurring (induced by the HRC detector) to the PSF, leading to a closer match between model and data for on-axis sources than would be the case otherwise. 

We would like to highlight two more subtle points: First, the PSF image is projected on a grid of sky pixels that exactly matches the pixel grid of the actual observation. Therefore, the ``true" source position (\ie~the positional input into {\em ChaRT}) is not located exactly at the central pixel of the PSF image, but slightly offset from it by a sub-pixel margin. This offset is an effect which we later correct for by adding its value to the fitted position of the source. If we ignore this effect, we find deviations between fits of the same source at different image bin sizes.

Second, when comparing the ``true" source position for the PSF of the on-axis (off-axis angle $<30''$) source with the location of the apparent centroid (or ``center of mass") of the simulated PSF, we find that they do not coincide as perfectly as one would expect. Instead, there is a systematic offset on the order of $0.1''$ which always appears to point in the same direction on the detector even for different simulated roll angles. This behavior is also observed when performing the entire ray-tracing simulation with {\em MARX} only. The {\em Chandra} Help Desk confirmed that this behavior is unexpected, and probably indicative of the achievable limit on  astrometric precision. 
The presence of subtle systematic effects at a sub-pixel level therefore must be considered with  much care in our analysis.

In the next step, we fit the PSF models to the individual images of \rx\ and A, B, C using {\em Sherpa}\footnote{\url{http://cxc.harvard.edu/sherpa/}}, a modelling and fitting package developed for {\em Chandra} \citep{Sherpa}. We follow the thread ``Accounting for PSF Effects in 2D Image Fitting", according to which we convolve the PSF Image (which is normalized to one) with a narrow Gaussian of fixed width, but free to vary in $(x,y)$-position and amplitude. Additionally, our model incorporates a small, spatially uniform background component across the relevant image area. The convolution with a Gaussian of finite width is necessary to perform meaningful interpolations between pixels so that non-integer position values are possible and the source position is not ``quantized" to the grid of image pixels. 

Due to the Poisson nature of the data, we use the fit statistic {\tt cstat}, an implementation of Poisson likelihood that can in principle be used similarly to $\chisq$ for model comparison, but regardless of the number of counts per pixel.\footnote{See \url{https://cxc.cfa.harvard.edu/sherpa/statistics/}}  Furthermore, we use the differential evolution algorithm implemented as {\tt moncar} for optimization. 
After performing the fit, we use the methods {\tt conf} to get a rough estimate of fitting uncertainties, and {\tt reg\_proj} to obtain a precise view of the error contours (or equivalently the likelihood profile) in the $(x,y)$-plane. 

During the refinement of the fit parameters, we noticed that the reference point for the convolution does not seem to naturally coincide with what we specify as PSF {\tt center} parameter. 
This behavior is very similar to the one produced by the CIAO software bug reported in \citet{Becker12} which led to the extreme proper-motion velocity reported by \citet{Winkler07} and constitutes a potentially serious problem for the analysis of the fit results, since for off-axis sources, this can result in an offset from the best-fit position by as much as a few pixels. 
Therefore, in order to actually ``fixate" the center of the PSF, we additionally need to specify the hidden parameter {\tt origin} which makes the convolution process behave as expected.  
As an example, a complete impression of data and fitted models for a single observation (epoch 2010) is given in Figure \ref{PSFFits}. In Table \ref{FitResults}, we list all best-fit positions for the individual epochs. 

\begin{deluxetable*}{cccccrcc}[t!]
\tabletypesize{\scriptsize} 
\tablewidth{0pc}
\renewcommand{\arraystretch}{1.15}
\tablecaption{Optical and ``raw'' fitted X-ray positions and properties for all sources at all four epochs. \label{FitResults}}
\tablehead{ \\[-0.5pc]
{} & {} &  {}  &\multicolumn{2}{c}{X-ray}  & &  \multicolumn{2}{c}{Optical ({\em Gaia} DR2)}  \\ \vspace{-6pt} \\ 
\cline{4-5} \cline{7-8}  %\\ \vspace{3pt}
 ObsID       & Epoch  & Source & R.A. (J2000.)      & Decl. (J2000.)     & Counts & R.A. (J2000.) & Decl. (J2000.)        \\
 % ObsID       & Epoch  & Source & Right Ascension      & Declination            & Counts & Right Ascension & Declination        \\
   {}        &        &        & (h:m:s)       & (d:m:s)        &         & (h:m:s)  & (d:m:s)  }
\startdata                                                                          
   749       &1999.97 &  NS  & 08:21:57.4040(006)&$-$43:00:16.539(005)&3123  &      {}          &           {}        \\
             &        &   A  & 08:21:46.2906(049)&$-$43:02:03.308(132)&45&  08:21:46.2953(0)&$-$43:02:03.632(0) \\
             &        &   B  & 08:22:24.0205(146)&$-$42:57:59.362(109)&109 & 08:22:24.0049(0)&$-$42:57:59.393(0)   \\
             &        &   C  & 08:21:48.8703(103)&$-$43:01:28.104(180)&13& 08:21:48.8790(1)&$-$43:01:28.316(1)  \\
   4612      &2005.31 &  NS  & 08:21:57.3817(003)&$-$43:00:17.223(004)&6854&      {}          &           {}        \\
             &        &   A  & 08:21:46.3002(052)&$-$43:02:03.919(042)&121 & 08:21:46.2896(0)&$-$43:02:03.618(0)  \\
             &        &   B  & 08:22:24.0203(070)&$-$42:57:59.549(149)&178& 08:22:24.0047(0)&$-$42:57:59.348(0)  \\
             &        &   C  & 08:21:48.8849(130)&$-$43:01:28.597(178)&9 & 08:21:48.8542(0)&$-$43:01:28.280(0)   \\
11819/12201  &2010.61 &  NS  & 08:21:57.3262(002)&$-$43:00:17.463(005)&10490&      {}          &           {}        \\
             &        &   A  & 08:21:46.2617(036)&$-$43:02:04.089(063)&199& 08:21:46.2840(0)&$-$43:02:03.603(0)  \\
             &        &   B  & 08:22:23.9851(111)&$-$42:57:59.491(225)&170 & 08:22:24.0046(0)&$-$42:57:59.303(0)  \\
             &        &   C  & 08:21:48.8535(135)&$-$43:01:28.861(123)&19 & 08:21:48.8295(0)&$-$43:01:28.244(0)   \\
20741        &2019.09 &  NS  & 08:21:57.3078(003)&$-$43:00:17.017(004)&6208&      {}          &           {}        \\
             &        &   A  &08:21:46.3035(074)&$-$43:02:03.284(101)&81 & 08:21:46.2749(0)&$-$43:02:03.580(0)   \\
             &        &   B  &08:22:24.0607(339)&$-$42:57:58.841(193)&77 & 08:22:24.0043(0)&$-$42:57:59.231(0)   \\
             &        &   C  &08:21:48.8239(145)&$-$43:01:27.880(211)&18 & 08:21:48.7900(0)&$-$43:01:28.186(0)   \\
\enddata  
\tablecomments{The X-ray positions shown here correspond to the best-fit source position in sky coordinates returned by {\em Sherpa}, uncorrected for any astrometric offsets. The errors on X-ray source positions correspond to the maximum one-sided one-sigma error returned by the {\tt conf} task and are therefore just a crude estimate of the associated uncertainties. \textbf{The optical positions have been corrected for the proper motions of the respective calibrator stars.} For illustrative purposes, we also list rounded ``errors'' on the last digits of the optical positions, demonstrating that they are of very little importance for the overall uncertainty budget. The column ``Counts" lists the amplitude of the Gaussian model which was convolved with the PSF. This therefore corresponds to an estimate for the number of source counts with subtracted background.}
\end{deluxetable*}

There are several things to note about the fitting process: First, as expected, the fitting of two of the calibrator sources proves to be difficult, due to very high background emission (source B) and the quite limited photon statistics (source C). This makes the statistical errors on the astrometric calibration for these sources at least an order of magnitude larger than for the fit of the CCO position on the detector.
Second, there are systematic offsets between the best-fit X-ray positions and the known optical positions for the stars. This proves that an astrometric correction is justified and needed in order to obtain the highest possible precision on the final result. 
Also, we would like to highlight that the uncertainties represented by the fit contours cannot be described well with simple independent Gaussian errors in $x$ and $y$, since they show significant irregularities and interdependencies.

Finally, we observe small but significant deviations between our best-fit positions and the ones in \citet{Becker12}, most noticeably for the NS. This is probably related to differences between the actual PSF models fitted to the data, since it seems unlikely that the data itself are altered this drastically by our reprocessing. The differences between the two fits are found to be partly explainable by the deviations between nominal centroid and center of mass of the PSF model which we indicated earlier. At worst, this corresponds to a systematic error in the NS position of around $0.1''$, which could in principle severely bias our final proper motion estimate. 
However, any minor coordinate offset that is constant over the detector or scales only linearly with $x$ and $y$ will naturally be compensated by our coordinate transformation in Section \ref{Trafo}, since it would apply to all sources equally. 

\subsection{Transformation to the World Coordinate System}\label{Trafo}

\noindent In principle, there are many ways imaginable to align the coordinate systems of the individual observations. However, given the small number of calibrators, we attempt only two very common types of transformations to the WCS, similar to those applied in \citet{Winkler07}:

\begin{itemize}

\item \textbf{Translation:} For each observation, we determine an optimal transformation with two degrees of freedom, corresponding to a simple coordinate offset $(\Delta x, \Delta y)$ in an arbitrary direction:
\begin{equation}\label{Translation}
\small \left( \begin{array}{c}
x'\\
y'\\
\end{array} \right) = 
\left( \begin{array}{c}
x\\
y\\
\end{array} \right) +
\left( \begin{array}{c}
\Delta x\\
\Delta y\\
\end{array} \right) ,
\end{equation}
where $x$ and $y$ correspond to {\em Chandra} sky coordinates, and $x'$ and $y'$ represent WCS locations projected onto the sky coordinate system.

\item  \textbf{Scaling \& Rotation:} In addition to the simple translation, we allow for a small scale factor of the coordinate system $r$ and a rotation by a small angle $\theta$:
\begin{equation} \label{RotScale}
\small \left( \begin{array}{c}
x'\\
y'\\
\end{array} \right) = \left( \begin{array}{cc}
r \cos \theta & -r \sin\theta \\
r \sin \theta & r \cos \theta \\
\end{array} 
\right)  \left( \begin{array}{c}
x\\
y\\
\end{array} \right) +
 \left( \begin{array}{c}
\Delta x\\
\Delta y\\
\end{array} \right) .
\end{equation}
\end{itemize}

These two methods are analogous to the available modes of {\em CIAO}'s standard {\tt wcs\_match} script. For the determination of the optimal transformation parameters, we weight all three calibrators evenly. Thereby, their ``center of mass" is relatively close to the actual position of the NS (\ie~the location of the calibrators is not heavily biased towards a certain side of the detector). 

Given that a simple Gaussian description of the error is likely an oversimplification, we choose a slightly different approach than \citet{Becker12} to determine an absolute position of the NS: For each source, we take into account the values of the fit statistic on a finely spaced $(x,y)$-grid around the best fit, rather than propagating the best-fit and Gaussian uncertainties. The statistic values are extracted using the {\em Sherpa} task {\tt reg\_proj}. 
For each star $i$ ($i=A,B,C$), the ``C statistic'' $\mathcal{C}_{i}$ corresponds to the twice the negative logarithm of the Poissonian likelihood $\mathcal{L}_{i}$. Therefore, we can obtain probability values $P_{i}(x,y)$ for the position of the star at every point on the grid around the best-fit value by normalizing the total likelihood to one:
\begin{eqnarray}
P_{i}(x,y) &=& \frac{\mathcal{L}_{i}(x,y)}{\sum_{\hat{x},\hat{y}}\mathcal{L}_{i}(\hat{x},\hat{y})} \nonumber \\
        &=& \frac{\exp \left(-\frac{1}{2}\,\mathcal{C}_{i}(x,y)\right)}{\sum_{\hat{x},\hat{y}}\exp \left(-\frac{1}{2}\,\mathcal{C}_{i}(\hat{x},\hat{y}) \right)} ,
\end{eqnarray}
where we implicitly assume a flat prior over our $(x,y)$-grid, \ie~all viable (and realistic) fit locations are assumed to be within the range of our grid. 

From this, we can now propagate our fit uncertainties without making any strong assumptions on their shape. For the translation method, this is relatively straightforward if we space all our grid points evenly: For each star $i$, we take the differential between the {\em Gaia} location at $(x',y')$ and the coordinates of the probability contours at $(x,y)$ to obtain a distribution of translation vectors:
\begin{equation}
T_{i}(\Delta x,\Delta y)\, = \,P_{i}(x\!=\!x'\!-\!\Delta x,\,y\!=\!y'\!-\!\Delta y) .   
\end{equation}
We then average over the three stars by convolving these distributions (corresponding to a summation of the components) and dividing the resulting translation vector by 3\@. We convolve this average distribution with the distribution for the NS location $P_{\rm NS}(x,y)$ to obtain an estimate of its corrected WCS location.

For the scaling \& rotation method, we cannot use the same principle since a rotation will automatically ``mix" the $x$ and $y$ coordinates, so convolving them on a cartesian grid is not sensible. Instead we use the following numerical Monte Carlo technique: For each of the four objects (NS, A, B, C), we sample $N=10^6$ points (e.g. $x_{i,1}$, $x_{i,2}$,..., $x_{i,N}$) from their individual probability distributions $P_{i}(x,y)$. From the samples of A, B, C, we obtain a distribution of the four transformation parameters $\Delta x_{n}$, $\Delta y_{n}$, $r_{n}$ and $\theta_{n}$ by fitting them in Equation \ref{RotScale}. This corresponds to solving the following equation in a standard least-squares manner for each of the $N$ samples.
\begin{equation} \label{LeastSquare}
\left( 
\begin{array}{cccc}
x_A & -y_A & 1 & 0 \\
y_A &  x_A & 0 & 1 \\
x_B & -y_B & 1 & 0 \\
y_B &  x_B & 0 & 1 \\
x_C & -y_C & 1 & 0 \\
y_C &  x_C & 0 & 1 \\
\end{array} 
\right) \left( \begin{array}{c}
r \cos\theta\\
r \sin\theta\\
\Delta x\\
\Delta y\\
\end{array} \right) \approx \left( \begin{array}{c}
x'_A\\
y'_A\\
x'_B\\
y'_B\\
x'_C\\
y'_C\\
\end{array} \right) .
\end{equation}

We then apply the individual transformations as in Equation \ref{RotScale} to the simulated sample of neutron star locations to obtain the probability distribution for its absolute location. This method automatically provides us with an estimate for the most likely location of the CCO and detailed uncertainty contours, since it takes into account all likely positions of the individual calibrators. In contrast, we found that standard Gaussian error propagation of only the diagonal elements of the covariance matrix for the transformation parameters leads to an overestimation of the final error on $(x'_{\rm NS},y'_{\rm NS})$, since the transformation parameter values are strongly dependent on each other (\ie~there are large off-diagonal elements in the covariance matrix).

\begin{figure}[h!]
\centering
%\captionsetup{width=0.8\linewidth}
\includegraphics[width=1.0\linewidth]{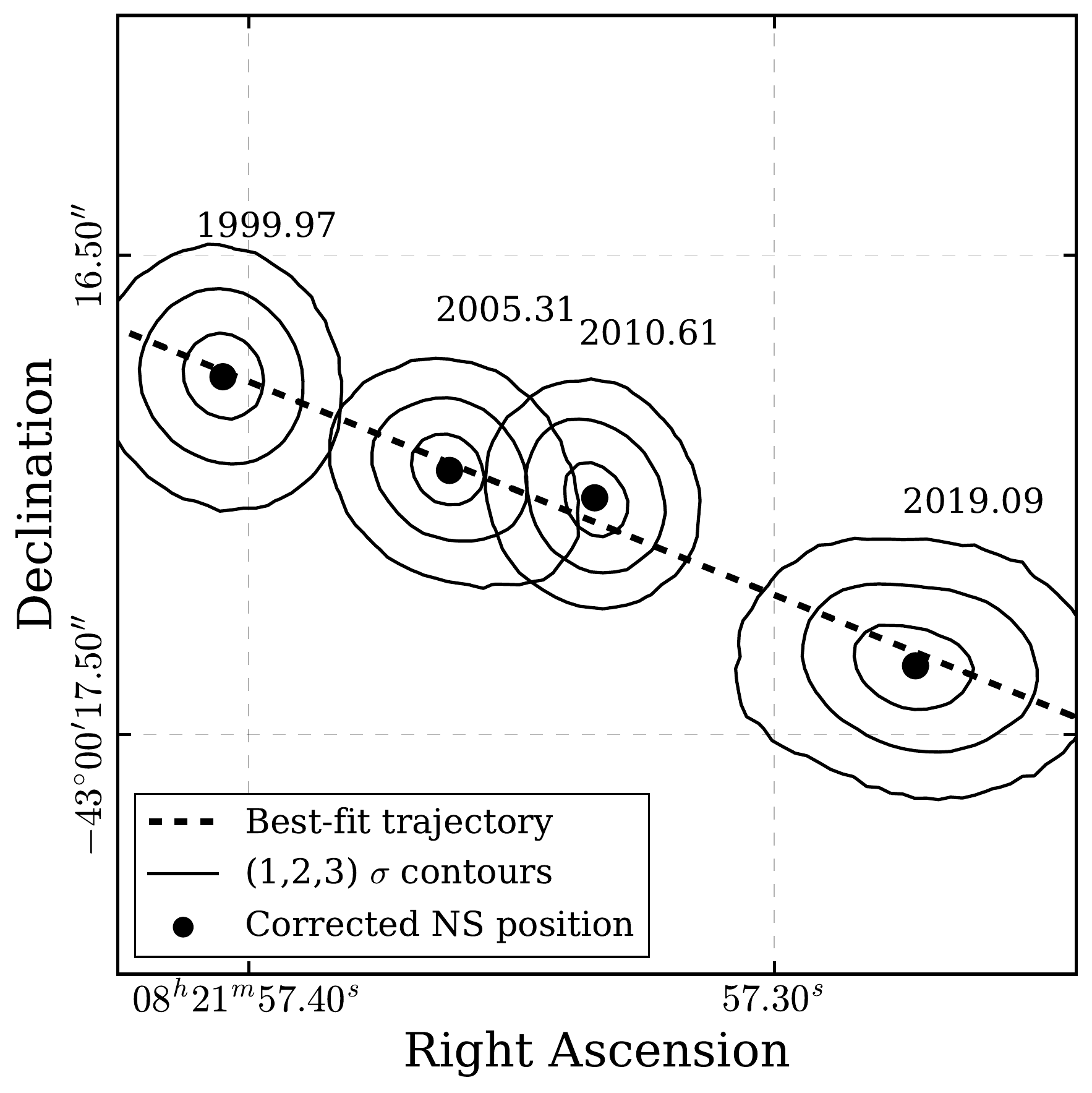}
\caption{Motion of \rx. We plot the absolute positions of the NS at the four epochs by indicating the mode and the 1, 2, and 3$\sigma$ contours (\ie~the smallest regions containing the corresponding fractions of cumulative probability) derived from their distribution. In addition, we indicate the direction of motion (\ie~the best-fit position angle $\phi_0$ as determined in Section \ref{LinearFit}) with a dashed line. This figure corresponds to an area of $2'' \times 2''$ on the sky.}
\label{PositionContours}
\end{figure}

By applying both methods to the PSF fits of each observation, and converting the resulting distributions from sky coordinates to celestial coordinates,\footnote{Here, the term ``sky coordinates'' refers to a tangent-plane system aligned with  celestial R.A. and Dec., but measured in {\em pixels}  (see \url{https://cxc.cfa.harvard.edu/ciao/ahelp/coords.html}).} we get a clear impression of the motion of the CCO, as can be seen in Figure \ref{PositionContours}. Note that here, as in the following sections, we choose to plot the results from the scaling \& rotation method, as it constitutes the more robust coordinate transformation, and its results barely differ from those from the translation method. The corresponding absolute positions, including uncertainties, for both methods are listed in Table \ref{CorrectedPos}. Here, as everywhere else in this paper, listed uncertainty ranges correspond to the $68\,\%$ central interval of the probability distribution of the respective quantity. Note that the relatively large uncertainty on the NS position in 2019 is caused by difficulties in the fitting of the position of source B. In that epoch, it is found to appear significantly fainter than in \eg, the observation from 2005, despite having comparable exposure times.

%\begin{figure*}[t!]

\begin{deluxetable*}{cccc}[t!]
\tabletypesize{\small}
\tablewidth{0pc}
\renewcommand{\arraystretch}{1.2}
\tablecaption{Absolute Positions of \rx \label{CorrectedPos}}
\tablehead{
%  Epoch &  Method & Right Ascension             & Declination                \\
  Epoch &  Method & R.A. (J2000.)             & Dec. (J2000.)                \\
    {}  &     {}        & (h:m:s)          & (d:m:s)            }
\startdata
1999.97 & Translation    &  08:21:57.402$^{+0.008}_{-0.008}$&$-$43:00:16.70$^{+0.07}_{-0.10}$ \\
        & Scaling \& rotation   & 08:21:57.405$^{+0.008}_{-0.008}$&$-$43:00:16.75$^{+0.09}_{-0.09}$  \\
2005.31 & Translation   &  08:21:57.360$^{+0.008}_{-0.007}$&$-$43:00:16.93$^{+0.08}_{-0.07}$ \\
        & Scaling \& rotation    & 08:21:57.362$^{+0.007}_{-0.007}$&$-$43:00:16.95$^{+0.07}_{-0.08}$ \\
2010.61 & Translation   & 08:21:57.331$^{+0.007}_{-0.008}$&$-$43:00:17.00$^{+0.07}_{-0.07}$\\
        & Scaling \& rotation & 08:21:57.334$^{+0.007}_{-0.006}$&$-$43:00:17.01$^{+0.08}_{-0.08}$\\
2019.09 & Translation   & 08:21:57.271$^{+0.014}_{-0.011}$&$-$43:00:17.33$^{+0.08}_{-0.09}$  \\
        & Scaling \& rotation & 08:21:57.273$^{+0.012}_{-0.011}$&$-$43:00:17.36$^{+0.09}_{-0.09}$ \\
\enddata  
\tablecomments{We list the median values and $68\,\%$ central intervals of the marginalized distributions for Right Ascension \& Declination at the given epochs.}
\end{deluxetable*}

\subsection{The Proper Motion of RX J0822-4300}\label{LinearFit}
\noindent From the probability distributions for the NS position at four epochs spanning 19.18 years, we can now determine the most likely value of its proper motion in a relatively straightforward way. We determine the best-fit values $\mu_{\alpha}$, $\mu_{\delta}$ fulfilling the following equation, describing motion at constant speed in two dimensions:  
\begin{equation}
 \left( \begin{array}{c}
\alpha(t)\\
\delta(t)\\
\end{array} \right) = 
\left( \begin{array}{c}
\mu_{\alpha}\\
\mu_{\delta}\\
\end{array} \right) \cdot (t-t_0)
+ 
\left( \begin{array}{c}
\alpha_0\\
\delta_0\\
\end{array} \right) ,
\end{equation}
where we have introduced the labels $\alpha$ and $\delta$ for Right Ascension and Declination, $t$ describes the epoch of the observations in years (with $t_0 = 2019.09$, corresponding to the time of our latest observation), and $\alpha_0$, $\delta_0$ correspond to the NS location at $t_0$. We define $\mu_{\alpha}$ such that a positive value describes an increase in Right Ascension, \ie~motion from west to east. 

\begin{figure}[h!]
\centering
%\captionsetup{width=0.8\linewidth}
\includegraphics[width=1.0\linewidth]{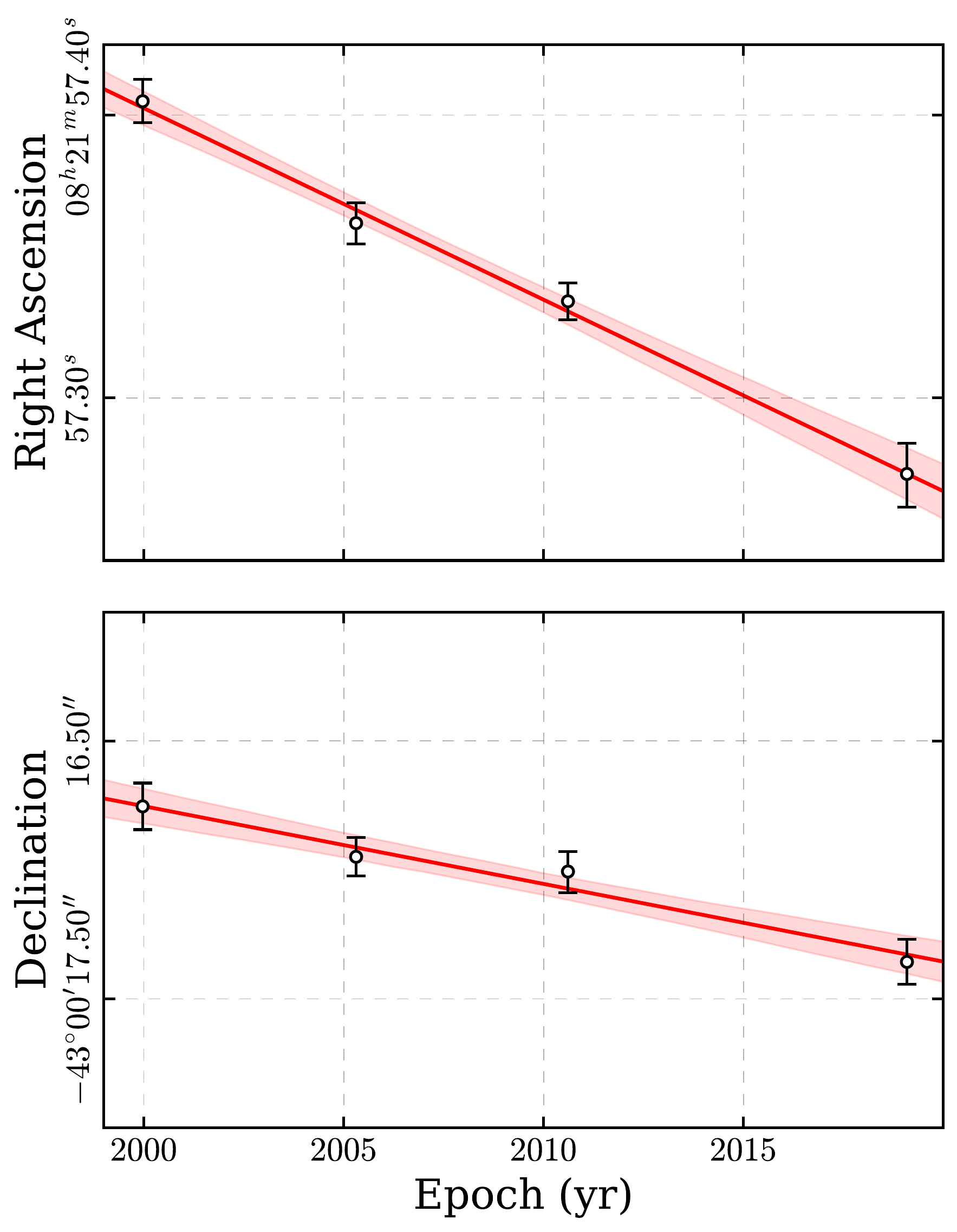}
\caption{Fits to the proper motion of \rx~ projected onto the right ascension (top) and declination (bottom) axes.  We indicate the median (best-fit) trajectory with a thick red line,  and the $68\%$ central interval of possible trajectories as red shaded regions.}
\label{LinFits}
\end{figure}

\begin{figure}[h!]
\centering
\includegraphics[width=1.0\linewidth]{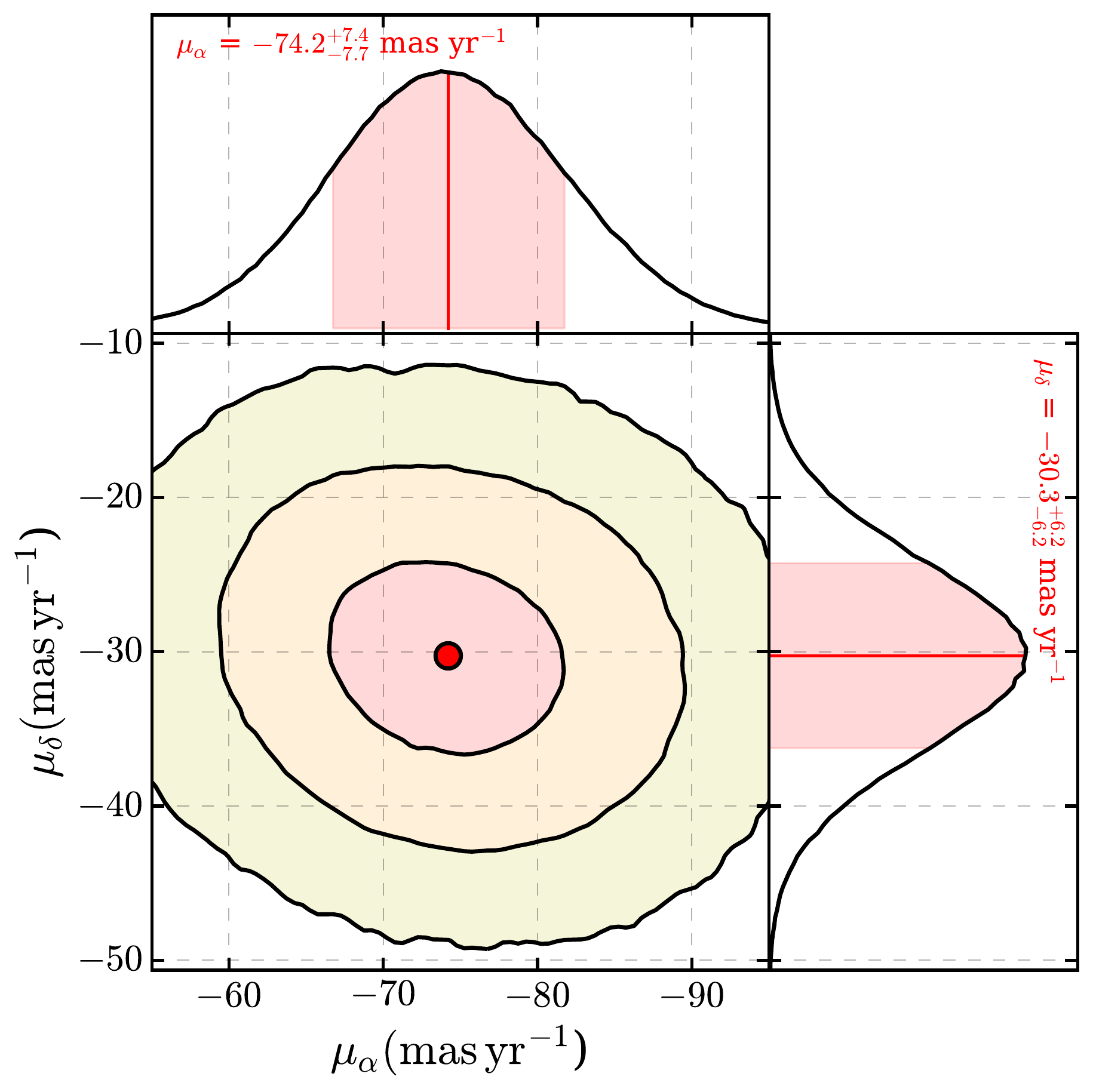}
\caption{2D distribution of the proper motion vector $(\mu_{\alpha}, \mu_{\delta})$. We show the best fit (red circle) and the contours corresponding to the cumulative probability within 1, 2, and 3$\sigma$, respectively. In the top \& right panels, we show the corresponding marginalized probability distributions for Right Ascension \& Declination components of proper motion. We indicate the median values \& $68 \%$ central intervals for the marginalized quantities in red.}
\label{RADECFits}
\end{figure}

In practice, we perform the fit by again drawing representative samples from the distributions for the individual epochs and then performing a least-squares fit for each sample, leading to a final distribution of proper motion values in $(\mu_{\alpha}, \mu_{\delta})$-space. With this method, we also obtain an absolute astrometric reference point $(\alpha_0, \delta_0)$ for \rx, corresponding to its position at the time of our latest observation (epoch 2019.09). By sampling simultaneously in $\alpha$ and $\delta$, we include the effect of any possible interdependence between these parameters, even though the position contours in Figure \ref{PositionContours} appear to be quite well behaved. We show representative one-dimensional projections onto the WCS axes of this fit (using the scaling \& rotation method) in Figure \ref{LinFits} and display the corresponding distribution of proper motion values in Figure \ref{RADECFits}.

%\begin{figure*}[t!]

%\begin{figure*}[t!]

The individual corrected positions at the four epochs agree well with the expected linear trajectory. Also, the probability distributions for the source locations and proper motion components appear well behaved and can be described with reasonable accuracy by Gaussian distributions. 

In order to exclude large systematic errors in our result due to a possibly biased PSF centroid (see Section \ref{PSFMod}), we also tried an alternative approach for the conversion of the fit results to the final proper motion value, by taking the ``center of mass'' of the PSF image as the precise source location instead of its nominal centroid position. From this analysis, we obtained results that  differ by only  $\sim 0.5 \masy$ from the ones shown here. This demonstrates that the effect of such minor potential offsets on the fit output can be balanced by our coordinate transformation method, which inherently compensates for linear distortions of the detector scale. 

\begin{figure}[h!]
\centering
%\captionsetup{width=0.7\linewidth}
\includegraphics[width=1.0\linewidth]{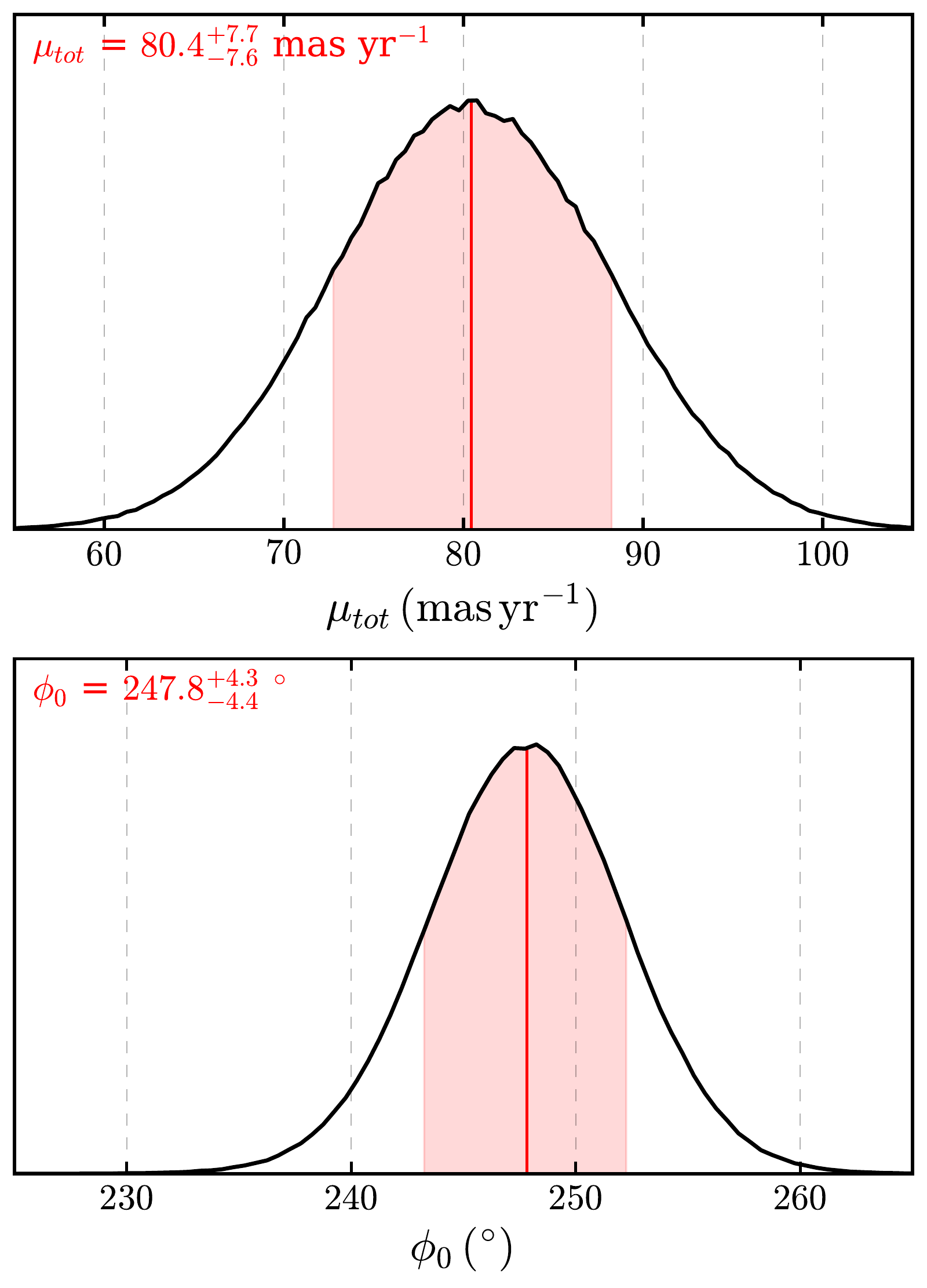}
\caption{{\em Top:} Marginalized distribution of the total proper motion $\mu_{\rm tot}$ with median and $68\,\%$ central interval indicated in red. {\em Bottom:} Same for the position angle $\phi_{0}$.}
\label{MUTOTPHI}
\end{figure}

In order to extract more illustrative quantities from our measurement, we convert the proper motion vector to polar coordinates by defining the total proper motion, $\mu_{\rm tot}$, and the position angle east of north, $\phi_0$, as:
\begin{eqnarray}
\mu_{\rm tot}\, &=&\, \sqrt{\mu_{\alpha}^2+ \mu_{\delta}^2\,} \\   
\tan \phi_0\, &=&\, \frac{\mu_{\alpha}}{\mu_{\delta}} .
\end{eqnarray}
By applying these simple relations to our sample of proper motion vectors, we obtain final probability distributions of the magnitude of the proper motion and of its direction, which we show in Figure \ref{MUTOTPHI}.

%\begin{figure*}[t!]

We display the resulting astrometric solutions and uncertainties in Table \ref{Results}. The results of the two transformation methods agree very well with each other. Since it constitutes the more robust coordinate transformation, we quote our proper motion from the scaling \& rotation method as final: $\mu_{\rm tot} = 80.4^{+7.7}_{-7.6} \masy$ and $\phi_0 = 247\fdg 8^{+4.3}_{-4.4}$. 

In general, our values agree within uncertainties with those given in \citet{Becker12} ($\mu_{\rm tot} = (71 \pm 12) \masy$, $\phi_0 = (244 \pm 11)^{\circ}$), with our median values corresponding to slightly higher proper motion and a slightly ``shallower'' position angle (when projected onto the sky). Naturally, our uncertainty on both values is smaller than theirs, since we have made use of a time baseline almost twice as long. Interestingly however, the relative increase in precision of the position angle is larger than that for the magnitude of proper motion. By looking at Figure \ref{PositionContours}, we can see that this is at least partly due to our position estimate for the observation from 2019, whose error contours are more extended along the direction of motion than perpendicular to it.

\citet{Gotthelf13} use an alternative method to constrain the proper motion of \rx: They determine the locations of the NS and calibrator stars by measuring their centroids of the observed images. They then correct the resulting value by the offset between centroid and source positions, as measured from simulated PSF images. They  use only star A as a calibrator source, since it is the brightest of the three stars. If we apply this method to our data set, we obtain a proper motion of $(82 \pm 7) \masy$ at a position angle of $(249 \pm 4)^{\circ}$, entirely consistent with our result from PSF fitting. 

While the statistical errors here are comparable with those from our method, the \citet{Gotthelf13} centroid method  uses only a single location on the detector to calibrate the astrometric reference frame, neglecting possible systematic distortions over the detector, which our scaling \& rotation method includes. It is possible to extend this "corrected centroid" method to stars B and C, but this is non-trivial as it requires an iterative procedure, and it leads to increased statistical errors. We conclude that while both methods are consistent, our PSF fitting technique is the more robust one.

\begin{deluxetable*}{ccccccc}[t!]
\tabletypesize{\small}
\tablewidth{0pc}
\renewcommand{\arraystretch}{1.2}
\tablecaption{Final Results for the proper motion of \rx \label{Results}}
\tablehead{
    Method & $\mu_{\alpha}$&$\mu_{\delta}$& $\mu_{\rm tot}$&$\phi_0$&$\alpha_0$&$\delta_0$ \\
           & $(\si{mas.yr^{-1}})$  & $(\si{mas.yr^{-1}})$ & $(\si{mas.yr^{-1}})$   &$(^{\circ})$ & (h:m:s) & (d:m:s)}
\startdata
Translation & $-75.1^{+7.7}_{-8.0}$ & $-31.0^{+6.4}_{-6.3}$     & $81.6^{+7.5}_{-7.5}$  & $247.5^{+4.7}_{-4.7}$ & 08:21:57.272$^{+0.009}_{-0.010}$ & $-$43:00:17.34$^{+0.08}_{-0.08}$\\
Scaling \& rotation & $-74.2^{+7.4}_{-7.7}$ & $-30.3^{+6.2}_{-6.2}$ & $80.4^{+7.7}_{-7.6}$ & $247.8^{+4.3}_{-4.4}$ & 08:21:57.274$^{+0.009}_{-0.010}$ & $-$43:00:17.33$^{+0.08}_{-0.08}$\\
\enddata  
\tablecomments{The proper motion values in this table correspond to the medians and $68\,\%$ central intervals indicated in figures \ref{RADECFits} and \ref{MUTOTPHI}. We provide $(\alpha_0,\delta_0)$ as reference point for the absolute astrometric position of \rx~at the epoch of our latest observation (2019.09, MJD 58516.5)}
\end{deluxetable*} 
%Epoch: 2019.09 MJD_0= 58516.52997106481
% RotScale:
%RA_0 = 8:21:57.274 +- (0.0090286220384464804, -0.0095476825943842086)
%Dec_0 = -43:0:17.33 +- (0.0760593712788733, -0.076594485834139192)
% Translation: 
%RA_0 = 8:21:57.272 +- (0.0094335546395950803, -0.0099767559042403062)
%Dec_0 = -43:0:17.34 +- (0.079259055290108904, -0.079479195462859309)

\section{Discussion\label{Discussion}}

\subsection{Kinematics and Kick Mechanism}

\noindent Our refined measurement of the proper motion of \rx~ agrees well with the results of \citet{Becker12}, while providing smaller error bars on its magnitude and position angle. Calculating the projected velocity of the neutron star tangential to the line of sight, $v_{\rm proj}$, at an assumed distance $d$\footnote{For the sake of comparability with earlier publications on this topic, we adopt a distance $d=2\,\si{kpc}$ as reference scale.}, we obtain
\begin{equation}
   v_{\rm proj} = 763^{+73}_{-72}\times \left ( \frac{d}{2\,\si{kpc}} \right ) \kms.
\end{equation} 
This in principle constitutes a lower limit on the kick which the neutron star experienced during the supernova explosion, and therefore an important constraint on supernova models. Generally, the conclusions on the kinematics of the system outlined by \citet{Becker12} hold when considering our updated value. In dependence of neutron star mass $M_{\rm NS}$ and $d$, we obtain the following expressions for the tangential components (or lower limits) of momentum $p$ and kinetic energy $E_{\rm kin}$ carried by the neutron star:
\begin{eqnarray}
   p\,  &=& \, (2.12 \pm 0.20) \times 10^{41} \nonumber \\
   &\times& \left ( \frac{d}{2\,\si{kpc}} \right ) \left ( \frac{M_{\rm NS}}{1.4 \,M_\sun} \right ) \, {\rm g\,cm\,s^{-1}} \\
   E_{\rm kin}\,  &=& \, 8.1^{+1.6}_{-1.5} \times 10^{48} \nonumber \\ 
   &\times& \left ( \frac{d}{2\,\si{kpc}} \right )^2 \left ( \frac{M_{\rm NS}}{1.4 \,M_\sun} \right ) \, {\rm ergs} .
\end{eqnarray}
Assuming a neutron star of mass $1.4 \,M_\sun$ at a distance of $2\,\si{kpc}$, we obtain an estimate for the momentum of the CCO of $p = (2.12 \pm 0.20) \times 10^{41}\, {\rm g\,cm\,s^{-1}}$. This is consistent with the approximate momentum attributed to the ejecta, seen to be expanding towards the northeast as fast, optically emitting filaments \citep{Winkler85,Winkler07}. 
For the kinetic energy of the neutron star, we obtain $E_{\rm kin} = 8.1 ^{+1.6}_{-1.5} \times 10^{48}\, {\rm ergs}$, corresponding to a fraction $f \sim 0.8\,\%$ of the energy released in a canonical core-collapse supernova explosion of $10^{51}\, {\rm ergs}$. 

While older measurements suggest a distance of around $2.2 \,\si{kpc}$ to Puppis A \citep[\eg][]{reynoso03}, several recent investigations favor a considerably lower distance of around $1.3\,\si{kpc}$ \citep{WoermannOHDistance,Aschenbach15,Reynoso17}. Assuming this lower distance would lead to a significantly smaller projected velocity of $\sim 500 \kms$, in even better agreement with the upper end of the neutron star velocity  distribution \citep[see \eg][]{hobbs05}. Furthermore, the inferred momentum and kinetic energy would be reduced accordingly to around $p \sim 1.4 \times 10^{41}\, {\rm g\,cm\,s^{-1}}$ and $E_{\rm kin} \sim 3.4 \times 10^{48}\, {\rm ergs}$, respectively. Distance measurements to Galactic supernova remnants are inherently difficult, since most  are based on measuring \ion{H}{1} or OH absorption features in their (continuum) radio spectrum, and using the presence (or absence) of such features, together with Galactic rotation models, to place lower and upper limits on the distance.
Alternative methods based on optical and/or X-ray absorption are typically at least as uncertain.   

In principle, natal kicks on neutron stars can occur \eg~via asymmetric neutrino emission during the explosion or via asymmetric ejection of matter due to hydrodynamic instabilities. The latter scenario is supported by the observed relationship between total ejecta mass and neutron star kick velocity \citep{BrayElridge}. \citet{Wongwa13} coined the term ``gravitational tug-boat mechanism'' for the underlying hydrodynamic mechanism: Massive, slowly moving ejecta on the side opposite the most violent explosion exert a gravitational pull on the newly born neutron star. This results in possible kick velocities on the order of $\sim 1000 \kms$ for strongly asymmetric explosions \citep{Janka17}. Therefore, our proper motion estimate for \rx, and the associated projected velocity are consistent with theoretical considerations for any reasonable assumption on the distance. 

The hydrodynamic nature of the kick mechanism is supported by an investigation of the spin properties of \rx: While the CCO does exhibit pulsed emission at a period of $0.112 \,\si{s}$, its origin is likely to be purely thermal, resulting from periodic modulation of black-body emission from two antipodal hotspots on the neutron star surface \citep{gotthelf09,gotthelf10}. The specific properties of these hotspots (temperature and effective area) lead to a phase-reversal of the pulse profile at an energy of around $1.2 \,\si{keV}$, rendering the broad-band detection of pulsed emission difficult. Through the analysis of phase-coherent timing observations, \citet{Gotthelf13} were able to measure a total period derivative of $\dot{P} = (9.28 \pm 0.36) \times\, 10^{-18}$ for the pulses of \rx. After consideration of the kinematic contribution of the neutron star motion via the Shklovskii effect \citep{Shklovskii}, they derived a magnetic field of around $2.9\times10^{10}\,\si{G}$ and a ``spin-down age'' corresponding to $\sim 2.5\times\,10^8 \,\si{yr}$. 

The latter quantity is, of course, an unrealistic age estimate, which shows that the implicit assumption of the neutron star being born rotating much faster than today is wrong for this object. In conjunction with the very weak magnetic field they inferred, this contradicts electromagnetic powering of the kick mechanism. Such would require the newly born neutron star to rotate very fast or exhibit a very large magnetic field \citep{lai01}.
The low magnetic field and small period derivative are shared with other members of the CCO class, thus justifying their designation as ``anti-magnetars'' \citep{Gotthelf13}. A possible explanation for the weak observed dipole field could be that it has been buried by rapid fallback accretion of supernova ejecta after the explosion, and only slowly diffuses back to the surface on a time scale of around $10^4\, \si{yr}$ \citep{Bogdan14,Luo15}.

\subsection{Age of Puppis A}
\begin{figure}[t!]
\centering
\includegraphics[width=1.0\linewidth]{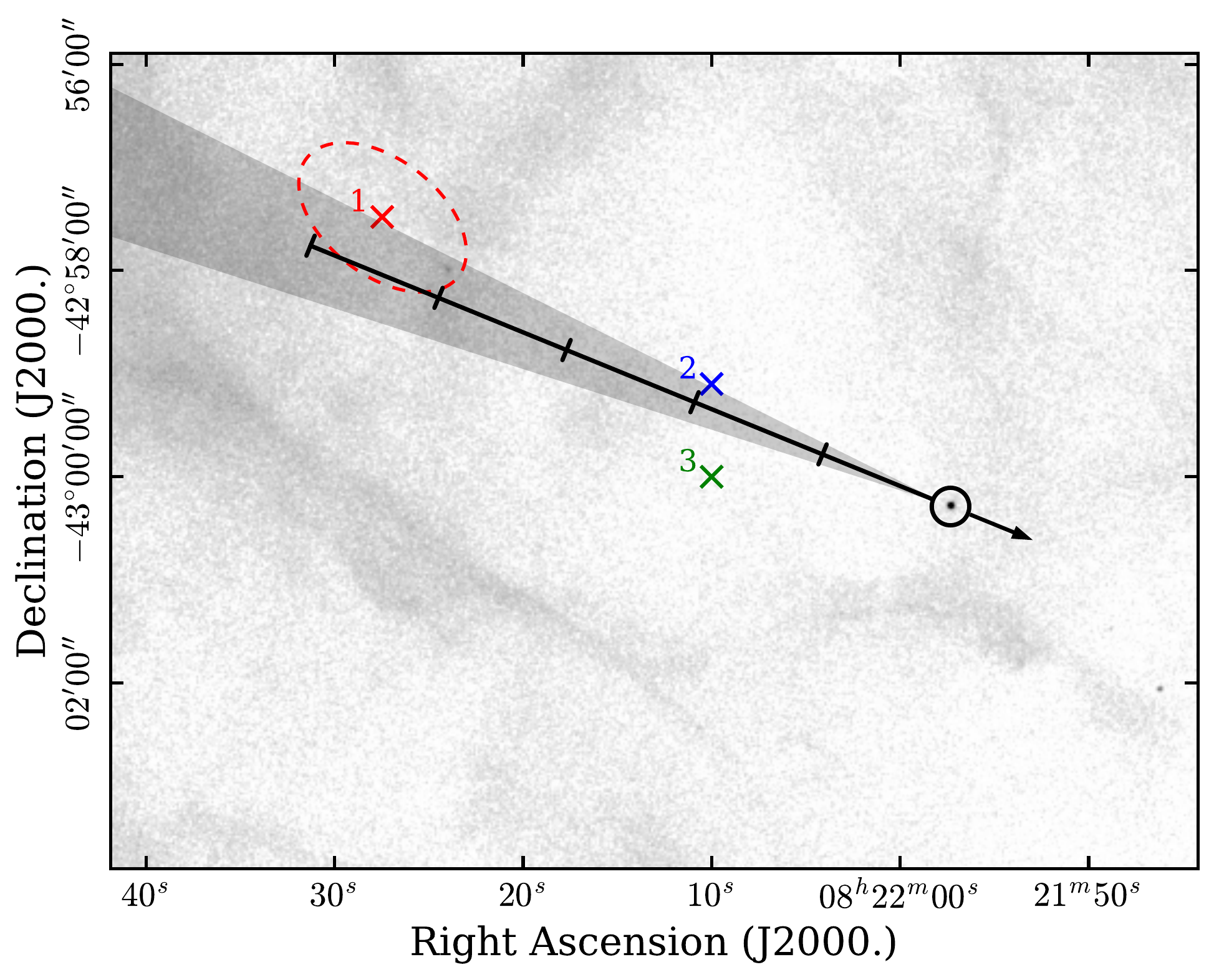}
\caption{X-ray image of the central region of Puppis A with the past trajectory of \rx~and directional uncertainties indicated. The optical expansion center of \cite{Winkler88} and its $68\,\%$ confidence ellipse is indicated in red and marked as 1. We also show the location of the alternative remnant center provided by \citet{Aschenbach15} and the remnant center obtained from radio data \citep[cf.][]{GreenCat} which are marked as 2 and 3, respectively. For \rx, we mark the distance travelled every 1000 years with increments.} 
\label{OptExpCen}
\end{figure}

\noindent By extrapolating the motion of the neutron star back in time, our revised proper motion measurement of \rx~also provides an updated estimate for the age of Puppis A. \citet{Winkler88} analyzed the motion of faint, oxygen-rich filaments of the SNR in the optical. They found expansion at very high velocities (up to $1500 \kms$) from a common center located at $\alpha(J2000) = 08^{h}22^{m}27.5^{s}$, 
$\delta(J2000) = -42^{\circ}57^{\prime}29^{\prime\prime}$. The semi-major axis of the $68\,\%$ confidence ellipse on this position is oriented almost exactly along the line toward the CCO (position angle $\phi = 242^{\circ}$ east of north), and measures $ 56\arcsec$; the semi-minor axis is $34\arcsec$ in the transverse direction.\footnote{The $68\%$ error ellipse comes from an updated analysis of the original data, and is (naturally) smaller than the $90\%$-confidence ellipse shown in \citet{Winkler88}.} 
Under the assumption of undecelerated trajectories for these dense knots, this center can be considered an estimate for the supernova explosion site. Assuming the errors on the expansion center to be approximately Gaussian, and comparing the coordinates of the expansion center with the position of the CCO in 2019, ($\alpha \approx 08^{h}21^{m}57.3^{s}$, $\delta \approx -43^{\circ}00^{\prime}17^{\prime\prime}$, Table~\ref{Results}), we find that the neutron star is located at an angular distance of $372\arcsec \pm 37 \arcsec$ from the expansion center determined by \citet{Winkler88}. The inferred direction of motion is $243^{\circ} \pm 4^{\circ}$, which overlaps, within the errors, with the position angle we measured for the proper motion of \rx~in X-rays. 
We illustrate its past trajectory and the location of the optical expansion center in Figure \ref{OptExpCen}.  

By weighting our sample of trajectories (Figure \ref{RADECFits}), according to their likelihood of overlap with the observed expansion center and determining the amount of time needed for the CCO to cover the observed angular distance given the respective proper motion, we obtain an estimate for the kinematic age $\tau$ of Puppis A:
\begin{equation}
    \tau = 4.6^{+0.7}_{-0.6}\times 10^3 \,\si{yr} . 
\end{equation}
This value is somewhat greater than the SNR age inferred from motion of the optical filaments alone, which \citet{Winkler88}  found to be $(3.7\pm0.3)\times 10^3  \,\si{yr}$, though the two values agree within the errors. The errors in the total displacement and position angle of the NS from its origin, and in the age of the SNR, are dominated by uncertainties in the expansion center for the system of ejecta filaments.

The neutron star itself is unlikely to have experienced any past deceleration, while the optically visible ejecta might have, due to their far lower density. Therefore, including a uniform deceleration model for the ejecta could possibly increase the minor tension between the two measurements, since the age inferred from optical filaments alone would then be reduced. As \citet{Winkler88} already noted, the apparent center of the radio shell is offset from the optical expansion center by $\sim 4'$ towards the southwest. Therefore, it may be worthwhile to consider that the actual explosion site might be located closer to \rx~than inferred, which would lead to a lower measured age from neutron star proper motion. 

\citet{Aschenbach15} proposes to include an ejecta deceleration model that is not radially symmetric, but allows for different degrees of deceleration along two perpendicular axes.
Repeating his approach with our updated proper motion value, the inferred age would be radically reduced by a factor $\sim 2.4$ to around $ 1950\,\si{yr}$.\footnote{\citet{Aschenbach15} originally states an age of $(1990 \pm 150)\,\si{yr}$, based on the proper motion of \citet{Becker12}, which enters his calculations explicitly.}  
The implied location of the remnant center would then be at $\alpha = 08^{h}22^{m}10.0^{s}$, $\delta = -42^{\circ}59^{\prime}06^{\prime\prime}$, lying within one arcminute of the center of the radio shell of the SNR as given in the Green catalogue \citep{GreenCat}. 
While the exact methodology may be a matter of debate here, this example highlights how strongly the kinematic age estimate can be systematically affected by input assumptions, such as an assumed (or neglected) deceleration model.

\subsection{Proper Motion Measurements of other CCOs}

\noindent Proper motion studies of neutron stars are generally a powerful tool for inferring their origin and age as well as the kinematics of the supernova explosion. For radio pulsars, such measurements exist in large numbers, allowing for statistical studies of their distribution \citep[\eg][]{hobbs05}. However, for neutron stars without radio emission, particularly CCOs, there are few such measurements, due to the paucity of objects and the challenging nature of such measurements at other wavelengths.

Apart from the measurement here and in previous works for \rx, X-ray proper motion results have been reported for three other CCOs, all with much lower transverse velocities than that of \rx.
For 1E 1207.4$-$5209, located in the SNR PKS 1209$-$51/52, \citet{Halpern15} measured a relatively small proper motion of $(15 \pm 7) \masy$, $v_{\rm{proj}} < 180 \kms$ for a distance of $2\,\si{kpc}$. 
For CXO J232327.8$+$584842 in Cas A, \citet{Delaney13} measured a marginally significant projected velocity of $(390\pm400)\kms$ for a distance of $3.4\,\si{kpc}$, which corresponds to a proper motion of $(24 \pm 25) \masy$. Lastly, for the proper motion of CXO J085201.4-461753 in the Vela Jr.~SNR (G266.2$-$1.2), only a $3\sigma$ upper limit of $<300 \masy$, corresponding to $< 1400\kms$ for a distance  upper limit of $1\,\si{kpc}$, could be determined \citep{Mignani19}.
The latter two cases suffered from a lack of nearby calibrator sources, explaining their relatively large statistical errors. 

All this illustrates that in order to perform proper motion measurements of CCOs to similar precision as in this work, the temporal baseline covered by the data must be quite long, the object should be located relatively nearby, and there must be astrometric calibrator sources in the field of view.

\section{Summary\label{Summary}}
\noindent We have incorporated a new {\em Chandra} observation of the central region of Puppis A to perform the most precise proper motion measurement of \rx~to date. In particular, we have generalized the treatment of positional errors and used all available information from optical calibrator stars to obtain reliable position estimates and errors at all epochs. Our results are consistent within errors with those of \citet{Becker12}. We obtain a projected velocity of $763^{+73}_{-72}\kms$, for a distance of $2\,\si{kpc}$ to Puppis A. 
While this value lies on the upper end of the observed neutron star velocity distribution, it does not pose a challenge for theoretical supernova models, since such speeds are achievable with hydrodynamical kick mechanisms. If the actual distance to Puppis A is smaller, as recent measurements suggest, then the velocity will become proportionally smaller as well.

The direction of neutron star motion is consistent with the measurement of the supernova explosion site from optical filament expansion by \citet{Winkler88}. Our new measurement of the proper motion implies an age of $4600^{+700}_{-600} \,\si{yr}$ for the remnant, which is somewhat older than that  derived from proper motions of the optical filaments  alone.   
An important pillar for our age determination of Puppis A is the location of the optical center of expansion. The best currently available estimate is now over 30 years old and was based on digitization of photographic plates from three epochs over a total baseline of only 8 years.  An updated measurement of the proper motions for the ejecta filaments based on CCD images, ideally from several epochs over an extended baseline, is long overdue.  Images for such a measurement are in-hand, and the results will be reported separately (Winkler et al., in prep.). 
If one then finds a significant disagreement between the age based on the motion of optical filaments and that from extrapolation of the neutron star trajectory, this could point towards non-ballistic motion of  the  supernova ejecta clumps due to their interaction with the surrounding ISM.   

\acknowledgments
\noindent 
We acknowledge multiple contributions by Rob Petre to the early stages of the work reported here.  His history with \rx\ dates to the {\em Einstein} Observatory era, when he suspected the existence of a point source within Puppis A, even before its formal discovery in {\em ROSAT} data \citep{Petre82, petre96}. 
MM acknowledges support by the International Max-Planck Research School on Astrophysics at the Ludwig-Maximilians University, IMPRS.  PFW acknowledges support from the NSF through grant AST-1714281 and from NASA through grant GO6-17064C.
We acknowledge the use of the {\em Chandra} data archive and we are thankful for support by the {\em Chandra} helpdesk at various stages of our analysis. 
This work has made use of data from the European Space Agency (ESA) mission
{\it Gaia} (\url{https://www.cosmos.esa.int/gaia}), processed by the {\it Gaia}
Data Processing and Analysis Consortium (DPAC,
\url{https://www.cosmos.esa.int/web/gaia/dpac/consortium}). Funding for the DPAC
has been provided by national institutions, in particular the institutions
participating in the {\it Gaia} Multilateral Agreement.

\end{document}